\documentclass[ reprint,amsmath,amssymb, aps,prb,superscriptaddress,
showpacs, showkeys]{revtex4-1}

\usepackage{color}
\usepackage{graphicx}
\usepackage{dcolumn}
\usepackage{bm}
\usepackage{hyperref}
\hypersetup{bookmarksnumbered, pdfpagemode=UseOutlines, 
colorlinks=true, citecolor=blue, filecolor=blue, linkcolor=blue, urlcolor=blue}
\graphicspath{X:\paper}
\begin{document} 

\title{Spin Hall magnetoresistance in paramagnetic NdGaO$_3$}

\author{V. Eswara Phanindra}
\thanks{corresponding author}
\email{e.p.vallabhaneni@rug.nl}
\author{A. Das}
\author{J. J. L. van Rijn}
\author{S. Chen}
\author{B. J. van Wees}
\author{T. Banerjee}
\thanks{corresponding author}
\email{t.banerjee@rug.nl}
\affiliation{University of Groningen, Zernike Institute for Advanced Materials, 9747 AG Groningen, The Netherlands}

\date{\today}

\begin{abstract}
In recent years, spin Hall magnetoresistance (SMR) has emerged as an efficient way to probe the spontaneous magnetization state in ordered magnetic systems, by electrical current. Less known is its versatility as a probe of materials that do not possess spontaneous magnetization such as in paramagnets.  
%but it has been confined to systems with a spontaneous magnetization. 
%In this work, we report on the first observation of SMR in a paramagnet-NdGaO$_3$(NGO)-a rare earth oxide. 
In this work, SMR is used to probe paramagnetic NdGaO$_3$(NGO), a rare earth oxide, possessing a sizable spin orbit interaction (L=6). NGO has not been investigated earlier for its efficiency in propagating spins.
%and less is known about its magneto-crystalline anisotropy and its role on spin transport.  
We have performed extensive temperature and angle dependent-magnetoresistance (ADMR) studies along dissimilar crystallographic axes in NGO, using platinum (Pt) as spin injector and detector and utilizing (inverse) spin Hall effect. We find a close correlation between the temperature dependence of the ADMR response with magnetization in NGO and a linear current bias dependence of the ADMR amplitudes. These are chacteristics of SMR effect in Pt/NGO, arising from the torque acting on localized moments in NGO and considering crystal field induced intermultiplet transitions with temperature. %Further, the surface sensitivity of SMR enabled us in identifying the influence of the Pt/NGO interface quality on spin transport. 
Control experiments on Pt/SrTiO$_{3}$ and Pt/SiO$_2$ devices were also carried out in order to validate the observed SMR response in Pt/NGO bilayer and to rule out magnetoresistive contributions from Pt. 
%(Hanle magnetoresistance and weak anti-localization). Overall, this work depicts the isolated magnetic moments in paramagnetic systems suitable for spin transport as promising materials in spintronics.
\end{abstract}
\maketitle
%\keywords{Spin Hall magnetoresistance, paramagnets, magneto-crystalline anisotropy}

\section{Introduction}
Spin Hall magnetoresistance (SMR) is commonly used as an efficient means to electrically access the magnetic order of an underlying magnetic material, using normal metals (NM) such as Pt with large spin orbit coupling (SOC)\cite{Chen2013, Chen2016, Althammer2013,Nakayama2013}. SMR is primarily a resistance modulation effect, 
arising from a collective interaction due to   the spin Hall effect, spin-transfer torque, and inverse spin Hall effect at the (anti-) ferromagnetic insulator FMI or AFI/NM interface\cite{Vlietstra2013a,Chen2016,Isasa2016,Hoogeboom2017,Fischer2018,Baldrati2018,Lebrun2019,Phys2021}. Earlier spin transport and SMR studies focused mostly on ordered magnetic systems, exhibiting spontaneous magnetization, whereas studies on paramagnetic insulators (PM-I)  are scarce\cite{Oyanagi2019,Liang2018,Oyanagi2020a,Lammel2019}.  
However, study of SMR in paramagnets, materials that are confined to a passive role, for instance as spacer layers, in spin valves\cite{Parkin1990} might be useful for designing new spintronic devices.  
%Since paramagnetic insulators devoid of any stray fields or Barkhausen noise\cite{Kronmuller1983} typical for ordered magnets, thereby offers the higher bit packing density in memory devices. 
In order to realise and establish paramagnet-based spintronics, thorough characterization of these materials by spin transport based approaches such as SMR, is thus essential. Further, SMR measurements on PMs provide information about interfacial parameters such as spin conductance, which in turn govern the interfacial exchange interaction (J$_{exc}$) between the conduction electrons in the Pt layer and localized moments of the PM layer. Experimental evaluation of these parameters will be important in advancing the  microscopic theory of SMR\cite{Zhang2019}.\\
%helpful for designing novel spintronic devices with high spin injection efficiency.\\
\begin{figure}[ht]
	\centering
	\includegraphics[scale=2.2]{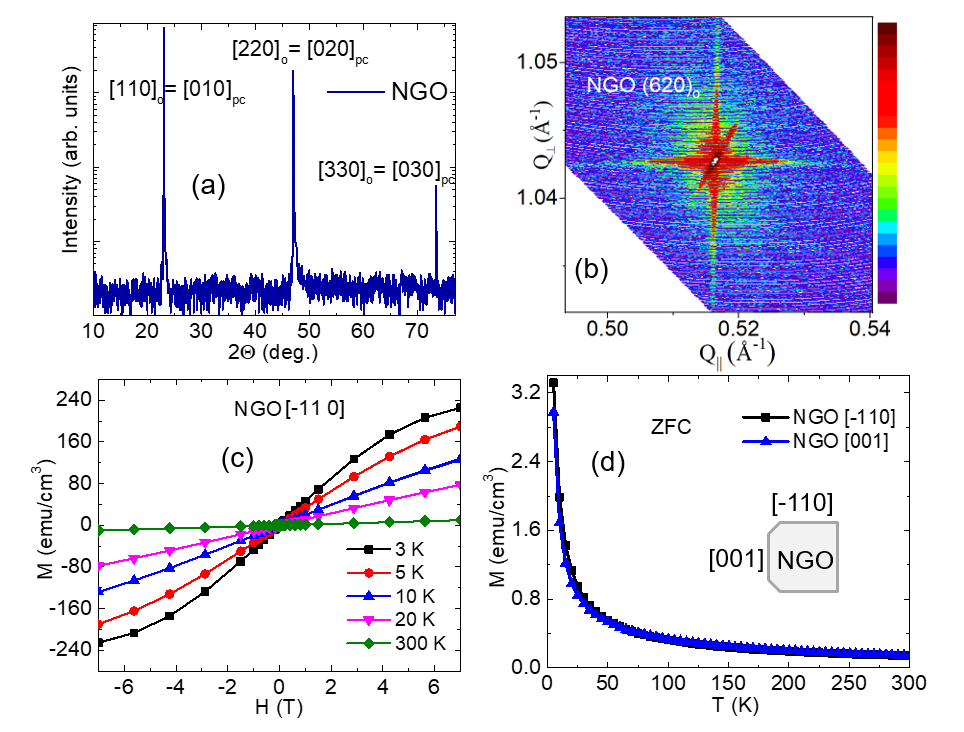}
	\caption{(a, b). $\theta-2\theta$ scan and reciprocal space map (RSM) of (6 2 0) asymmetric reflection of NdGaO$_{3}$ (110)$_{o}$ single crystalline substrate.  Fig. 1(c) Magnetization(M) vs applied magnetic field (H) loops measured at different temperatures.  Fig. 1(d) Magnetization(M) vs temperature (T) plots measured at an applied field of H =1000 Oe along two in-plane orthogonal (i.e. [001]$_{o}$  and  $[-110]_{o}$) crystallographic axes.  }
\end{figure} 
In this regard, we experimentally demonstrate the SMR effect in a new paramagnetic insulator, NdGaO$_{3}$ (NGO). NGO, has a good lattice match for the growth of high quality and strained complex oxide thin films such as rare earth manganites, high T$_{C}$ superconductors, etc\cite{Luis1998}. NGO exhibits an orthorhombic distorted perovskite structure with lattice constants denoted in orthorhombic notation as a = 5.43 Å, b = 5.70 Å, and c = 7.71 Å belonging to the Pbnm space group\cite{Boschker2009}, while in pseudo cubic notation the lattice constant of NGO corresponds to about 3.86 Å. For this study, we chose [110]$_{o}$ oriented NGO along the out-of-plane direction, with the in-plane edges correspond to  [-110]$_{o}$ and [001]$_{o}$ directions. Although NGO is a PM insulator, at temperatures below 1 K, it orders into an antiferromagnetic (AFM) phase, with dissimilar exchange (J$_{ex}$) interactions along the out-of-plane and in-plane directions\cite{Luis1998}. In DyScO$_{3}$ (PM-I), analogous to NGO, it was shown that short range correlations can favor the manifestation of spin Seebeck signals (SSE)\cite{Wu2015} at very low temperatures. Recent studies on PM-Gd$_3$Ga$_5$O$_{12}$ (GGG), using SMR, demonstrate efficient spin current transport persisting up to $\sim$ 100 K and ascribed to the dipolar exchange interactions between the localized spins\cite{Oyanagi2019}.  A theoretical study on SSE suggests that the paramagnetic-SSE signal is proportional to external magnetic field times the spin susceptibility of the magnet (or field-induced magnetization in PM-I)\cite{Yamamoto2019}. \\
%The current interest in PM materials motivated us to investigate the spintronic properties of NGO upto room temperature.\\
In this manuscript we perform SMR studies in paramagnetic-NGO single crystalline substrates and observe spin transport persisting up to $\sim$250 K, far beyond the $T_N$. The persistence of the observed SMR signals over a large temperature range suggest the inadequacy of the earlier explanations such as, dipolar exchange interactions \cite{Oyanagi2019} and short-range correlations, as applied for GGG\cite{Wu2015}, to explain our observations. 
%as in the case of GGG may not be apt due to observation of SMR signals over wide temperature range in our case. 
To understand our findings, we recall the magnetic and crystal field induced intermultiplet transitions with temperature of the Nd$^{3+}$ ions, as discussed in earlier studies on NGO\cite{WMarti,Podlesnyak1993, Novak2013}. We propose the self-interactions between the correlated 4f orbitals due to crystal field effects, J mixing between the multiplets, the temperature driven electronic excitations between crystal field split levels and spin flip scattering across HM/PI to explain our observations \cite{De2021,Podlesnyak1993}. This is distinctly different from those used to explain SMR effects in GGG/Pt  system  \cite{Wu2015, Oyanagi2019}.\\  
%to understand/explain the spin-transport across the HM/PI interfaces.\\
%Overall, this is a first study to probe spin transport in PM-NGO, and establish its utility in spintronics. Further, the surface sensitivity of the SMR technique enabled us in identifying the role of Pt/NGO interface quality on spin transport.
Our observation of SMR upto high temperatures ($\sim$ 250 K) in PM-NGO is  unique and implies that long-range magnetic ordering might not be an essential pre-requisite for efficient spin transport in paramagnets. 
%as probed by SMR %plemented by earlier reports on GGG\cite{Oyanagi2019}, 
Our work not only enriches material perspectives but triggers new considerations to the theoretical framework of SMR while encompassing material systems that do not exhibit spontaneous magnetic ordering. \\ 

\vspace{-2em}\section{Structural and Magnetic Characterization}
Crystal structure measurements such as $\theta$-2$\theta$ scans and reciprocal space maps(RSM’s) were performed on a NGO single crystalline substrate using PANalytical x-ray diffractometer equipped with a four-axis cradle (Cu k$_{\alpha}$  radiation, $\lambda$ = 1.54 Å) to substantiate the phase purity and its crystallographic directions [Fig. 1(a)]. $\theta$-2$\theta$ scan display only multiples of [110]$_o$ peaks along the out-of-plane direction, complemented by RSM scan [Fig. 1(b)], which depicts high substrate crystalline quality devoid of any twin domains or additional impurity phases. Similarly AFM micro-graphs within the scanning area of 5 × 5 $\mu$m$^2$  indicate smooth surface morphology of NGO with a surface roughness of 101 pm [Fig. S1]. Temperature (5-300 K) and magnetic field dependent (upto 7 T) magnetization curves of NGO substrate were measured by SQUID magnetometer (Quantum Design). \\
%respectively
\begin{figure*}[ht]
	\centering
	\includegraphics[scale=4.65]{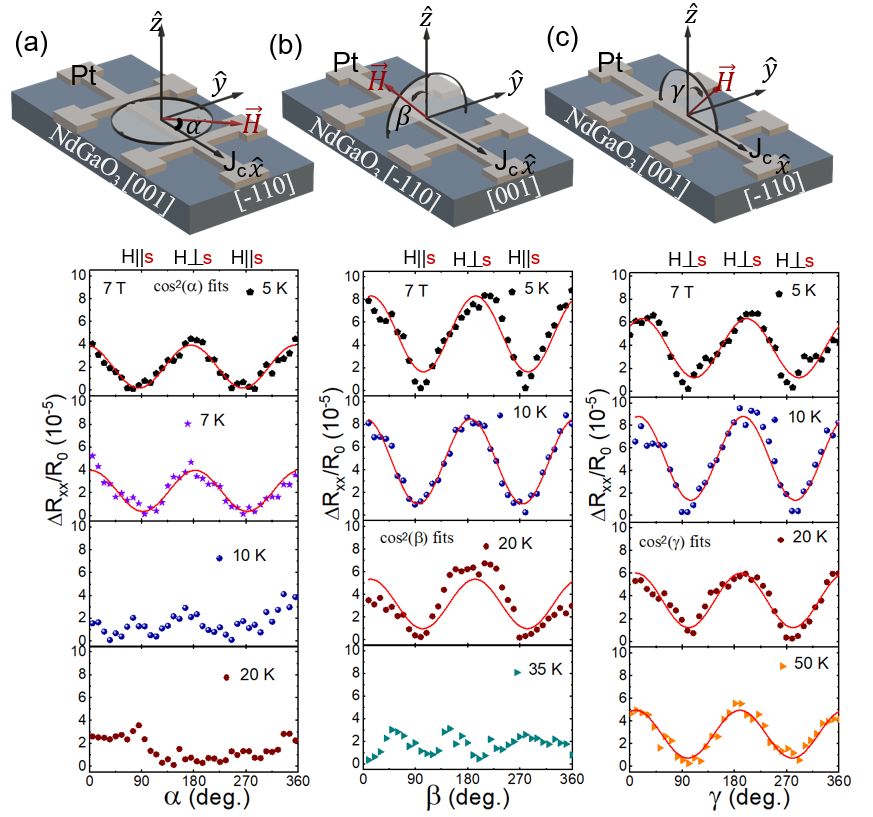}
	\caption{
Schematic illustration of the Hall bar geometry (top panel) and angle dependent magneto-resistance (ADMR) curves (bottom panels)  in the longitudinal contacts ($\Delta R_{xx}/ R_o$) in Pt (11 nm)/NdGaO$_3$ (110)$_{o}$, for selected temperatures where ADMR signals show noticeable changes. The ADMR signals are measured at different temperatures, and by rotating the sample with respect to a constant magnetic field of 7 T along (a) x-y ($\alpha$-scan), (b) y-z ($\beta$-scan) and (c) x-z ($\gamma$-scan)  planes.  Here s, J$_c$, H represents the corresponding spin accumulation, charge current and magnetic field directions respectively. The red solid lines are fits to experimental data with a cos$^{2}$($\alpha$) function.}
\end{figure*} 
%\textbf{Magnetization studies:}
Fig. 1(c) shows the Magnetization(M) vs applied magnetic field (H) loops measured at different temperatures.  Fig. 1(d) depicts the magnetization vs temperature (M–T) curves recorded in zero field cooled (ZFC) and field cooled (FC) modes from 300 K to 5 K at an applied field of 1 kOe along two in-plane orthogonal (   [001]$_{o}$ and [-110]$_{o}$) 
%asindicated by CrysTec GmbH) 
crystallographic axes for NGO sample. Both the M vs H and M vs T plots of the NGO substrate indicate a typical paramagnetic behavior, over a wide temperature range. At low temperatures, the difference of magnetization acquired along in-plane orthogonal (i.e. [001]$_{o}$  and [-110]$_{o}$) axes though subtle, is clearly visible, indicating [-110]$_{o}$  direction as the relatively easy axis and [001]$_{o}$ direction as the hard axis respectively. This difference is ascribed to the magneto-crystalline anisotropy in NGO\cite{Steenbeck2011}. 
\section{Magnetotransport measurements}
For magneto transport measurements, 11 nm thick-Pt based Hall bar devices (dimensions: width w = 30 µm, length l = 3000 µm) on NGO single crystalline substrate (as shown in Figs. 2, S2) are patterned and fabricated by photo-lithography followed by sputtering techniques respectively. An optical image of the Pt Hall bar device fabricated on top of NGO single crystalline substrate along two (i.e. [001]$_{o}$  and  $[-110]_{o}$) crystallographic directions is presented in the supplementary Fig. S2. All the angle dependent magneto-resistance (ADMR) measurements are performed on physical property measurement system (PPMS, Quantum Design) from 5 K to 300 K and upto 7 T magnetic field strength. An alternating sinusoidal charge current was applied to the Pt Hall bar and the voltage response across the longitudinal and transverse directions were simultaneously measured as a first harmonic response $V_{xx}^{1\omega}$  and  $V_{xy}^{1\omega}$ at phase = 0{\textdegree}  by two lock-in amplifiers. Typically, the ADMR measurements were performed with an AC current of amplitude 3 mA, at a reference frequency of 7.77 Hz and a time constant of 3 seconds was employed while rotating the sample in the in-plane(x-y) and out-of-plane(x-z), (y-z) configurations.\\
%\begin{figure*}[ht]
%	\centering
%	\includegraphics[scale=0.7]{FIG 3 RXX.PNG}
%	\caption{(a, b, c) depicts the magnetic field dependent magnetoresistance (FDMR) data ($\Delta \rho_{xx}/ \rho_0$) obtained along the longitudinal contacts for the three orthogonal crystallographic axes at 10 K, for Pt/NGO, Pt/SrTiO$_3$ and Pt/SiO$_2$ devices respectively. The corresponding measurement schematic illustration (black, red, blue curves corresponds to $\hat {x}$, $\hat {y}$, $\hat {z}$ directions respectively) employed for acquisition of FDMR plots.   }
%\end{figure*}
%\textbf{Spin transport measurements at Pt/NGO interface:}
Fig. 2(top panel) shows the schematic illustration of the angle dependent measurement geometry and the corresponding sample rotation along the x-y ($\alpha$), y-z ($\beta$) and x-z ($\gamma$) directions with respect to the applied magnetic field direction. The electrical current density (J$_c$) along the x direction in Pt layer induces spin accumulation ($\sigma$) along y and spin current density (J$_s$)  along z-direction (due to spin Hall effect), thus mutually orthogonal to each other and related as \cite{Chen2013} 
%${J_s = \theta_{SH}(\sigma \times J_c)}$. 
The spin current density (J$_s$) induced in the Pt layer exerts a torque on the field induced net magnetization (M) vector of the underlying NGO substrate. Depending on the relative orientation between M and {$\sigma$}, reflection or absorption of spin current occurs at the Pt/NGO interface which in turn causes the modulation of spin dependent magneto-resistance. Thus the resistance maximizes when ${M{\perp}\sigma}$  %M\bot${\sigma}$,
while minimal resistance change occurs for the parallel case (i.e. M// $\sigma$)\cite{Althammer2013}.  The typical longitudinal and transverse SMR response at the Pt/NGO interface can be expressed %modelled 
as follows\cite{Chen2013,Chen2016,Fischer2018}:

\begin{equation}
\frac{\Delta R_{xx}}{R_o} = \frac{\Delta \rho_{xx}}{\rho_o} <1-m_y^{2}> \propto cos^2(\alpha) 
\end{equation}
\begin{equation}
\frac{l}{w}\frac{\Delta R_{xy}}{R_o} = 2\frac{\Delta \rho_{xy}}{\rho_o}<m_x m_y>  \propto  sin(2\alpha)  
\end{equation}

where $\Delta R_{xx}/ R_o$ and $\Delta R_{xy}/ R_o$ are the longitudinal and transverse SMR amplitude variations respectively. Similarly, m$_{x}$, m$_{y}$ and m$_{z}$ denotes the projection of the net magnetization along the three orthogonal axes. R$_o$ is the Drude resistance, l = 325 \textmu{m} is the distance between the longitudinal contacts, w = 30 \textmu{m} is the width of Hall bar respectively. Fig. 2(a, b, c) shows the ADMR curves in Pt/NGO sample along three directions as shown in the schematic illustration at low temperatures for 7 T field. The longitudinal magnetoresistance variations shown for Pt/NGO, follows a $cos^2(\alpha)$ variation implying positive SMR response, consistent with the equations (1,2), as usually observed in ferro-, ferri-, and para-magnets\cite{Oyanagi2020a,Althammer2013,Chen2016}. \\
\begin{figure*}[ht]
	\centering
	\includegraphics[scale=3.68]{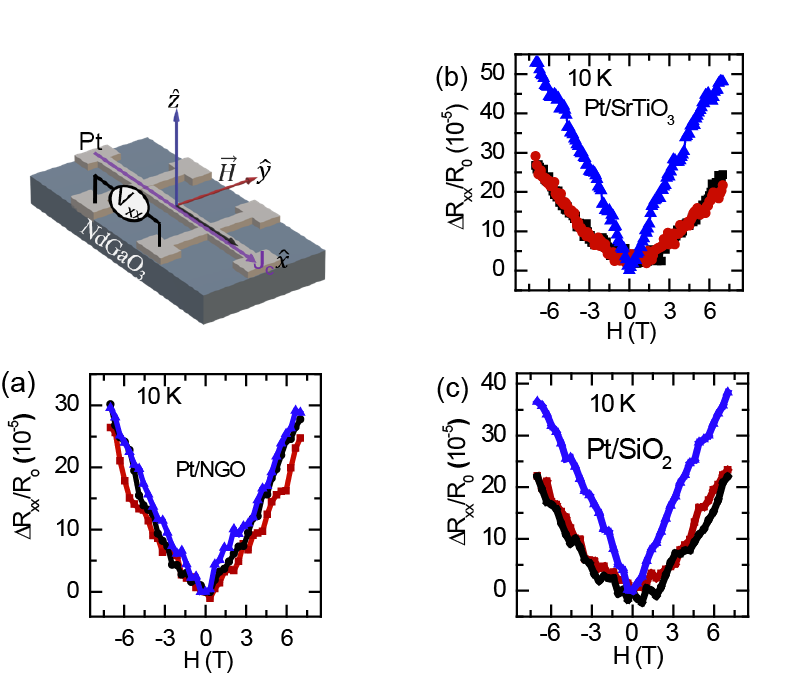}
	\caption{(a, b, c) Magnetic field dependent magnetoresistance (FDMR) data ($\Delta \rho_{xx}/ \rho_0$) obtained along the longitudinal contacts for the three orthogonal crystallographic axes at 10 K, for (a) Pt/NGO, (b) Pt/SrTiO$_3$ and (c) Pt/SiO$_2$ devices respectively. The corresponding measurement schematic illustration (black, red, blue curves corresponds to $\hat {x}$, $\hat {y}$, $\hat {z}$ directions respectively) employed for acquisition of FDMR.   }
\end{figure*}

Additionally, Fig. 2(c) depicts resistance modulation in the $\gamma$ scan at low temperatures, which can arise due to weak antilocalization (WAL)\cite{Velez2016a}, and is usually observed in heavy metals at low temperatures and high magnetic fields. Moreover for both $\beta$ and $\gamma$ scans (out-of-plane scans, Fig. 2(b, c)) the magnitude of the resistance modulation is larger than the in-plane $\alpha$ scan and typical of WAL. To confirm this and to  decouple it from SMR, we have performed a thorough and systematic field dependent magnetoresistance (FDMR) measurements on Pt/NGO, Pt/SiO$_{2}$ and Pt on single crystalline cubic SrTiO$_{3}$ devices, along three orthogonal directions as shown in  Fig. 3. Along the out-of-plane direction (Fig. 3), all the devices exhibit qualitatively similar field dependent signature of WAL (independent of the underlying substrate), implying the intrinsic contribution of Pt and consistent with previous reports\cite{Velez2016a, wu2016b}. Further, the FDMR response (Fig. 3) measured along the in-plane directions for both the Pt/SiO$_{2}$ and Pt/SrTiO$_{3}$ devices show a quadratic MR response characteristic of a positive-ordinary MR reiterating the intrinsic contribution of the Pt layer\cite{Velez2016a, wu2016b}. On the other hand the MR response for the Pt/NGO device, exhibits deviation from the quadratic behavior, implying a minor contribution due to SMR. Thus the FDMR measurements shown in Fig. 3, performed on control devices with different para- and dia-magnetic substrates, effectively explains our observed ADMR signals in $\alpha$, $\beta$ and $\gamma$ scans.\\
Other contributions, such as Hanle magnetoresistance (HMR), can be present, in addition to SMR, in the observed resistance modulation along the $\alpha$ and $\beta$ directions as shown in Fig. 2a, b. HMR, typically ascribed to the dephasing of the spin accumulation at the Pt interface, exists in pure Pt independent of adjacent magnetic layer and scales with the applied field strength\cite{Velez2016a, wu2016b}. In contrast, the SMR requires the presence of an  adjacent para-/(Anti)ferro-/ferri-magnetic layer. In order to disentangle the HMR response from SMR, control experiments on a Pt/SiO$_{2}$ device, were carried out in order to confirm the observed SMR response in Pt/NGO bilayer (Figs. 4 [d, e]).
As is seen in Fig. 4(d), angle dependent magnetoresistance is barely observed in the Pt/SiO$_{2}$ device, implying that the observed ADMR response in NGO along in-plane configuration results from SMR effect. This supports the fact that the NGO moments are indeed responsible for the observed ADMR response along the in-plane direction. Also the temperature dependence of ADMR amplitudes obtained for both Pt/NGO and Pt/SiO$_{2}$ exhibits distinct temperature dependent behavior, characteristic of SMR and HMR\cite{Velez2016a, wu2016b} respectively[Fig. 4 (d, e)].  \\
\begin{figure*}[ht]
	\centering
	\includegraphics[scale=0.482]{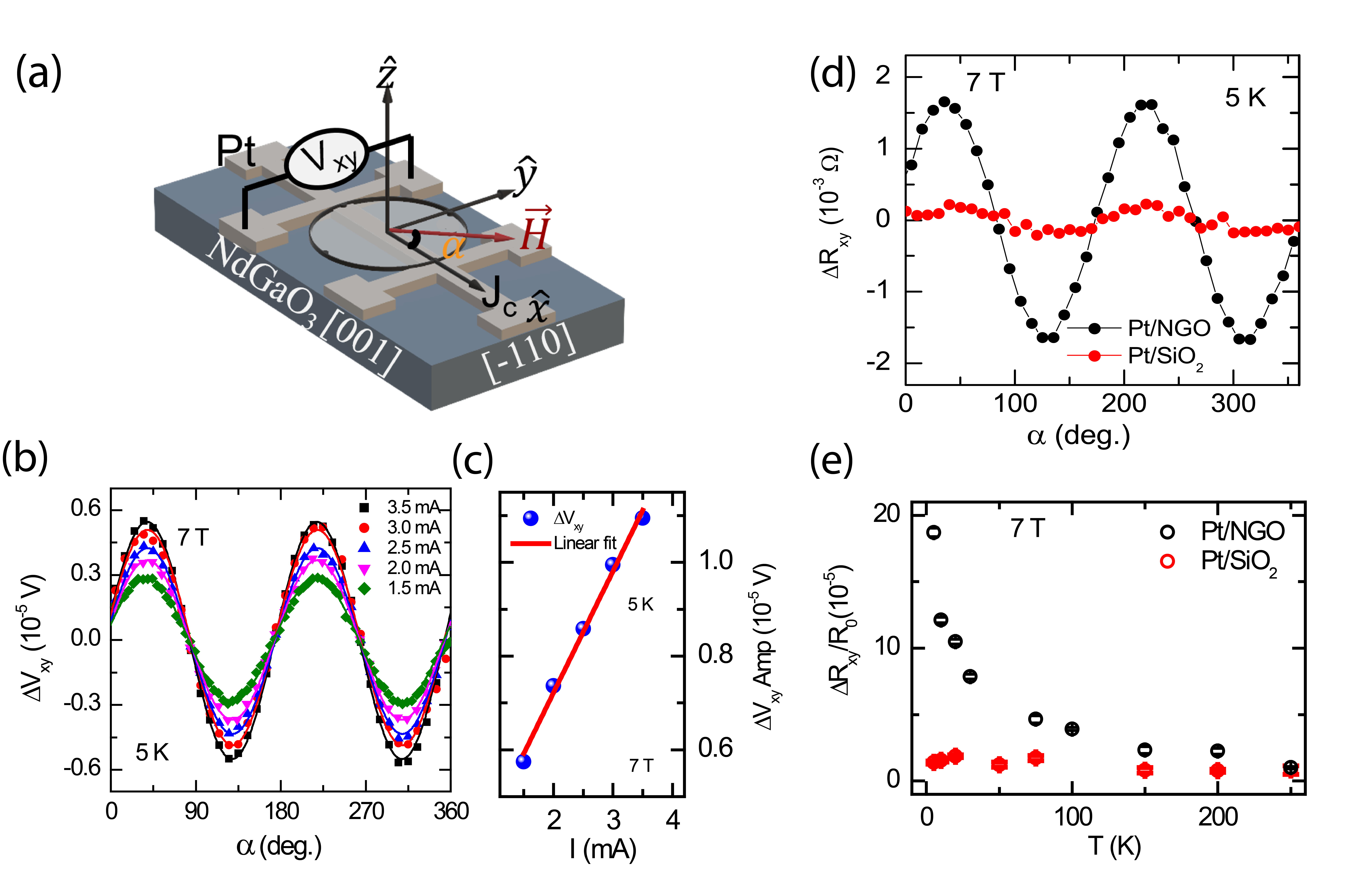}
	\caption{(a) Schematic illustration of the Hall bar geometry employed for observation of transverse-SMR response across  (11nm) Pt /NdGaO$_3$ (110)o. (b) Angular dependent magnetoresistance (ADMR) signals (V$_{xy}$) are shown for different current bias, and by rotating the sample in the plane of the transport (x-y plane) at a constant magnetic field of 7 T. (c) shows the linear current dependence of $\Delta V_{xy}$ amplitude obtained from the $sin2(\alpha)$ function fits (solid lines) to experimental data ($\Delta V_{xy}$) at 5 K. (d) shows the comparison of ADMR signals obtained for Pt/NGO and Pt/SiO$_{2}$ samples acquired for lowest possible temperature 5 K and high magnetic field of 7 T in our system in order to observe any SMR signals for Pt/SiO$_{2}$. As one could infer even for 5 K and 7 T field, Pt/SiO$_{2}$ sample exhibits negligible ADMR response in comparison to Pt/NGO, (e) shows the temperature dependence of ADMR amplitudes obtained for Pt/NGO and Pt/SiO$_{2}$ samples at 7 T field which clearly depicts two distinct behaviors i.e. SMR response in Pt/NGO and characterstic HMR response (as described in the text) in Pt/SiO$_{2}$.}
\end{figure*}
%\textbf{Temperature dependence of transverse-SMR ($\Delta R_{xy}/ R_o$) signals:} 
The observation of a small longitudinal-SMR response (i.e. $\Delta R_{xx}$/R$_{o}$, as observed in ADMR measurements) and its reduction with temperature (shown in Fig. 2a) can be attributed to a large Drude background resistivity (resulting from a 11 nm-thick Pt layer) which significantly masks the longitudinal SMR response\cite{Aqeel2015}. Thus, to minimize the background resistance influence of the Pt layer, we have performed measurements in the transverse configuration (i.e. $\Delta R_{xy}/ R_o$). The obtained (ADMR) response in the transverse configuration at 5 K displays a clear sin(2${\alpha}$) feature, consistent with Eq. 2 (Fig. 4b). Also, the SMR amplitudes acquired by fitting the $\Delta V_{xy}^{1\omega}$ signal to sin(2${\alpha}$) function scales linearly with the applied current bias implying that the measured ADMR response originates from the SMR effect\cite{Fischer2018}[Fig. 4c]. \\
\begin{figure*}[ht]
	\centering
	\includegraphics[scale=3.46]{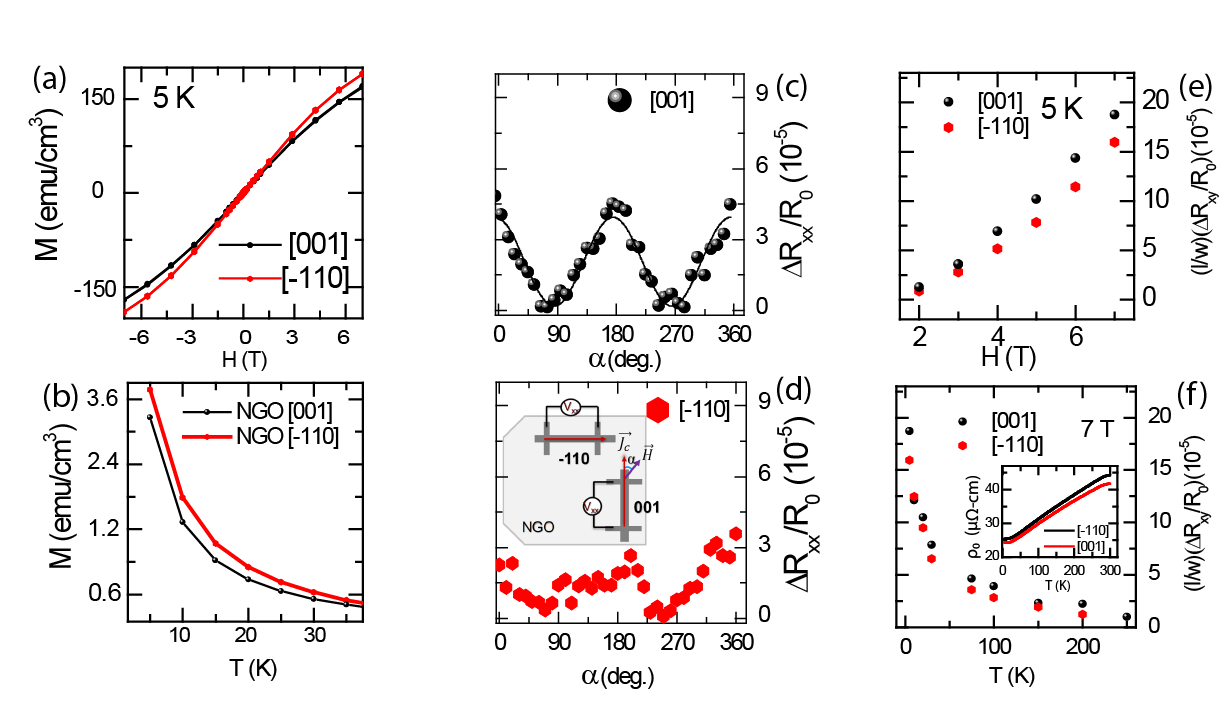}
	\caption{(a, b) Magnetization (M) vs applied magnetic field (H), and M vs temperature (T) plotted along two in-plane orthogonal (i.e. [001]$_{o}$  and  $[-110]_{o}$) crystallographic axes. (c, d) Angular dependent spin Hall magnetoresistance (ADMR) signals ($\Delta R_{xx}$/$R_{0}$) measured along the longitudinal contacts across (11 nm) Pt /NdGaO$_{3}$ (110)$_{o}$ sample by rotating the sample in-plane at a constant field of 7 T, and by applying the current bias along (001) and (-110) directions respectively (solid line is a fit to SMR data following eq. 1).  (e, f) depicts the magnetic field and temperature dependence of $(\frac{l}{w})(\frac{\Delta R_{xy}}{R_o})$ amplitudes respectively along two in-plane axes(here l,w are length, width of longitudinal contacts as per eq. 1). Inset to 5 (f) depicts the corresponding temperature dependent longitudinal Drude resistivity of Pt along the two in-plane axes. }
\end{figure*}
%\textbf{Influence of Pt/NGO interface quality on transverse SMR signals:} 
Further to the analysis of the obtained data, we consider the role of magnetic anisotropy in such orthorhombic substrates. Orthorhombic substrates such as DyScO$_3$\cite{Ke2009}, NdGaO$_3$\cite{Luis1998}, etc. are known to exhibit magnetic anisotropy at low temperatures along dissimilar crystal axes since anisotropy in magnetic properties or susceptibility can occur generally in any crystalline structure which deviates from the cubic symmetry. This is manifested as differences in the magnetization data acquired along [001]$_{o}$  and  [-110]$_{o}$ axes in NGO and are the relatively hard and easy axes respectively[Fig. 5(a, b)]. In order to study the impact of magnetic anisotropy on spin transport we have fabricated Pt Hall bars oriented with respect to the in-plane crystalline axes of the NGO substrate (shown in Fig. 5d). This allows us to investigate the SMR response for two different current directions i.e. along the  [-110]$_{o}$ direction and the [001]$_{o}$  directions of NGO respectively. Such studies can provide useful insights on the role of magnetic anisotropy and Pt/NGO interface quality on spin transport. \\
From equations 1 and 2, it is evident that any contribution from the magnetic anisotropy will be prominently manifested in the SMR response measured along the longitudinal configuration (longitudinal SMR response is dependent only on the m$_y$ component, i.e. $\Delta R_{xx}/R_o  \propto  <1-m_y^2>  \propto cos^2 (\alpha)$). Thus, a larger longitudinal SMR response is expected along the relatively easy axis i.e.  [-110]$_{o}$ compared to [001]$_{o}$  axis (according to the magnetization data Fig. 5[a, b]). However, we find a finite longitudinal SMR response along [001]$_{o}$ axis (relatively hard axis) but negligible SMR response along [-110]$_{o}$ (relatively easy axis) at low temperatures [Fig. 5(c, d)].\\
On the other hand, the SMR response measured along the transverse direction should be insensitive to the magnetic anisotropy of the underlying magnetic layer according to the eq. 2, since $\Delta R_{xy}/R_o  \propto  <m_x m_y>  \propto sin(2\alpha)$. Thus, the transverse SMR response measured along the orthogonal in-plane axes should yield similar ($\Delta R_{xy}/R_o$) amplitudes, which is contrary to our experimental results along  [-110]$_{o}$ and [001]$_{o}$  directions [Fig. 5(e, f)]. Such distinct transverse SMR response may be attributed to subtle differences in the NGO surface morphology, either due to substrate steps and terraces, evident in AFM images or to variations in the Pt thickness and morphology, or related to the fabrication of different Pt Hall bars on the orthorhombically distorted-NGO substrate. The observation of different Drude resistivity values of Pt along orthogonal axes [Fig. 5f(inset)] is in line with our argument above and consistent with the previous report\cite{Althammer2019}.\\
%Fig. S5 explains  
The minor differences in the transverse SMR results (Fig. 5(e, f)) and Pt resistivity values along  [-110]$_{o}$ and [001]$_{o}$ directions [Fig. 5f(inset)] can be explained 
%as follows: a lower Pt resistivity along [001]$_{o}$ axes implies a smoother Pt/NGO interface with 
by reduced interface scattering and incremental enhancement in the effective spin conductance\cite{Althammer2019}. 
%due to a smoother surface morphology along the [001]$_{o}$ axis 
This can facilitate interaction between the 6s spins in the Pt layer and Nd$^{3+}$ 4f orbital spins across the Pt/NGO interface, leading to the observation of a larger transverse SMR amplitude along [001]$_{o}$ as compared to [-110]$_{o}$ axis.\\
Further, we also find that the temperature dependence of the SMR amplitude decreases and vanishes beyond 250 K for the Pt/NGO device at 7 T as shown in Fig. 5(f). This follows the temperature dependence of the magnetization in NGO. The temperature dependence of SMR in NGO and its reduction at high temperatures can be correlated to the increased thermal agitation of the localized spins. This highlights SMR as an effective probe of magnetization in materials that not only exhibit long range ordering but also  that arises from localized spins such as in PM.\\ 
%of ordered magnets but also for PM materials such as NGO. The temperature dependence of SMR in NGO and its reduction at high temperatures can be correlated to the increased thermal agitation of the localized spins. \\ %outbalancing the Zeeman energy.\\
%(which [$\Delta$R$_{xy}$/R$_o$] otherwise has to be identical along both the axes). \
%\textbf{Longitudinal SMR-influence of Pt/NGO interface quality:}
\\
%\textbf{Spin current transport across Pt/NGO interface:}

\section{Discussion}

The interaction between the Nd$^{3+}$ moments in paramagnetic NGO is not strong enough to yield the observed SMR signals existing upto high temperatures. In GGG, with weak spin orbit interaction, the observed spin transport was explained in terms of dipolar exchange interactions \cite{Oyanagi2019} or short-range correlations \cite{Wu2015} whereas in another paramagnet, Tb$_3$Ga$_5$O$_{12}$, with a finite orbital angular moment (L=3), the authors do not observe any such signatures. They highlight the importance of the weak spin-lattice coupling in long-range paramagnetic spin transmission in GGG. In NGO with L=6, a strong correlation between spin and orbital angular momentum is expected, however we do observe clear signatures of SMR. To understand the underlying mechanism, 
%governing the spin transport across the Pt/NGO interface which inturn leads to the manifestation of SMR signals (also the weak interaction between the Nd$^{3+}$ localized magnetic moments). Here 
we propose the role of temperature driven electronic excitations between the crystal field split levels (ground [4I$_{9/2}$] and excited states[4I$_{11/2}$]) in Nd$^{3+}$. In related studies on neutron diffraction in NGO \cite{Podlesnyak1993}, the authors unambiguously ascertain four inelastic peaks of magnetic origin and show that the ten-fold degeneracy of the ground-state [4I$_{9/2}$] of the Nd$^{3+}$ ions is split by the crystalline electric field (CEF) into five Kramers doublet levels. At temperatures between 12-150 K, the thermal energy (k$_{B}$T) drives the transitions between these doublet ground state [4I$_{9/2}$] levels with the energy spectra lying between 11 and
70 meV. Further increasing temperature (beyond 150 K) leads to the additional electronic excitations/transitions between the excited states [4I$_{11/2}$]) in Nd$^{3+}$. This sets in self-interaction between 4f orbitals in NGO, leading to ordering of localized moments.\\ 
%Spin flip scattering (at HM/PI) \cite{Oyanagi2019} induced spin excitations between the Zeeman split levels (due to external magnetic field i.e. $g_{j} \Delta m_j\mu_B B, \Delta m_j= \pm 1$ represents the possible allowed transitions).\\
%In the first case, according to Podlesnyak et. al., the ten-fold degeneracy of the ground-state [4I$_{9/2}$] of the Nd$^{3+}$ ions is split by the crystalline electric field (CEF) into five Kramers doublet levels. At low temperatures (i.e. from 2-150 K), thermal energy (k$_{B}$T) drives the transitions between these doublet ground state [4I$_{9/2}$] levels which varies in the range of 70 meV. Further increasing temperature (beyond 150 K) leads to the additional electronic excitations/transitions between the excited states [4I$_{11/2}$]) in Nd$^{3+}$.\\
%Next, in addition to these thermal driven electronic excitations between the crystal field split levels, the external magnetic field further splits the five doublet levels, and the spin flip scattering mechanism (at HM/PI) induce non-equilibrium state of localized spins over wide temperature range as follows: 
We also consider interfacial spin flip scattering \cite{Oyanagi2019} induced spin excitations between the Zeeman split levels (i.e. $g_{j} \Delta m_j\mu_B B, \Delta m_j= \pm 1$ represents the possible allowed transitions, here g$_{j}$ is Lande g factor, m$_j$ is magnetic quantum number, $\mu_B$ is the Bohr magneton). In presence of an external magnetic field, the degeneracy of Nd$^{3+}$ localized 4f orbitals is lifted further, leading to different energy levels (as shown in Fig. S3a) where the energy spacing (i.e. $g_{j} \Delta m_j\mu_B$) scales with the external field strength. At Pt/NGO interface, the spin interaction between Pt (6s) conduction electrons with Nd$^{3+}$ localized spins alters the spin state by an energy of $g_{j} \Delta m_j\mu_B B$ leading to a non-equilibrium state (Fig. S3b) of localized spins. 
%(similar to spin flip scattering at HM/FM interface where it creates and absorbs magnon). 
This results in a gradient of the spin chemical potential ($\mu_{PI}$) and a spin current diffusion that is proportional to the difference in spin chemical potential $\mu_{PI}$ and electronic chemical potential $\mu_{HM}$, across the interface\cite{Oyanagi2019} as follows:
\begin{equation}
    J_s = (\frac{h}{2\pi e^2}) g_s (\mu_{PI} - \mu_{HM})
\end{equation}
Here g$_s$ is the effective spin conductance which governs the coupling between the spin chemical potential and the spin accumulation across the Pt/NGO interface.
\section{Conclusion}
Thus we conclude that that the magnetic properties in perovskite paramagnetic NGO, that leads to the observation of SMR upto ~250 K has its origin in the self-interactions between the correlated 4f orbitals due to crystal field effects, J mixing between the multiplets and the temperature driven electronic excitations between crystal field split levels.  Our angle dependent-MR measurements reveal spin dependent magnetoresistance modulation effect in  paramagnetic NGO persisting beyond T$_{N}$ in a temperature regime  much higher than reported for paramagnetic GGG. %Further, the surface sensitivity of the SMR technique enables us to decouple the influence of the Pt/NGO surface morphology and the magneto-crystalline anisotropy on spin transport. 
Control experiments on Pt/SiO$_2$ sample were also carried out in order to substantiate the observed SMR response in Pt/NGO bilayer and decouple the role of  possible intrinsic effects in Pt due to HMR/WAL. Low temperature SMR studies (spanning across the NGO magnetic phase transition) will be interesting to understand the impact of AFM exchange interaction on the SMR response. Our work also underpins the need for new considerations to the theoretical framework of SMR in encompassing  diverse materials beyond those that order spontaneously and which exhibit non-zero  correlation between spin and orbital angular momentum.\\ %not only with spontaneous magnetization, but for  paramagnets where spin orbit interaction in materials with non zero orbital magnetic moment can also lead to magnetization ordering as revealed by SMR studies. 
%is p.  Furthermore, theoretical studies are needed to understand the role of the effective spin mixing conductance, s-f exchange interaction (between 6s spins in Pt layer and NGO 4f orbital spins) in such perovskite paramagnets with non zero orbital magnetic moment.\\
\section{Data Availability}
The data that support the findings of this study are available from the corresponding author upon reasonable request.\\
\section{Acknowledgements}
The authors thank A. S. Goossens, Jacob Baas, J.G. Holstein, H.H. de Vries, T. J. Schouten for technical support. This work is realized using the facilities available at NanoLab NL and is supported by the Dieptestrategie grant from the Zernike Institute for Advanced Materials, University of Groningen. V.E.P thanks financial support from the Spinoza Prize awarded to B. J. van Wees by NWO.

\bibliography{Pt-NGO_BIB.bib}
\end{document}